# Self-Organization Promotes the Evolution of Cooperation with Cultural Propagation

Luis Enrique Cortés-Berrueco[1], Carlos Gershenson[2,4], and Christopher R. Stephens[3,4]

[1] Graduate Program in Computer Science and Engineering, Universidad Nacional Autónoma de México, Ciudad Universitaria , A.P.70-600, México D.F., México
`lecortesb@comunidad.unam.mx`
[2] Computer Science Department, Instituto de Investigaciones en Matemáticas Aplicadas y en Sistemas, Universidad Nacional Autónoma de México, Ciudad Universitaria, A.P.20-726, México D.F., México
`cgg@unam.mx`
[3] Gravitation Department. Instituto de Ciencias Nucleares, Universidad Nacional Autónoma de México, Ciudad Universitaria, A.P.70-543, México D.F., México
`stephens@nucleares.unam.mx`
[4] Centro de Ciencias de la Complejidad,
Universidad Nacional Autónoma de México

**Abstract.** In this paper three computational models for the study of the evolution of cooperation under cultural propagation are studied: Kin Selection, Direct Reciprocity and Indirect Reciprocity. Two analyzes are reported, one comparing their behavior between them and a second one identifying the impact that different parameters have in the model dynamics. The results of these analyzes illustrate how game transitions may occur depending of some parameters within the models and also explain how agents adapt to these transitions by individually choosing their attachment to a cooperative attitude. These parameters regulate how cooperation can self-organize under different circumstances. The emergence of the evolution of cooperation as a result of the agent's adapting processes is also discussed.

## 1 Introduction

Urban traffic problems have a complex behavior even in their most simplistic abstractions [2]. This behavior becomes more complex when driver interaction is added [10]. With these two premises, in this paper, three computational models, which were originally intended for studying driving behaviors and their impact in traffic, are presented. The aim is to explore the possibilities for controlling some of the complex characteristics of traffic, e.g. the consequences of driver interactions. The possibility of using them for other social problems is also discussed.

The models were conceived to study the evolution of cooperation [1], which studies how cooperative behaviors are preserved when selfish behaviors offer greater individual rewards, e.g. drivers who yield at lane changes or crossings, animal females taking care of their offspring, students who don't betray each other. Several abstractions for simulating different circumstances in which this phenomenon occurs have been proposed [6]. The models are based in three abstractions of different observed situations in which cooperation evolves [4]. Table 1 shows these abstractions as theoretical games in which players may choose between two behaviors: cooperative or selfish. When a player chooses to cooperate she has to pay a cost that her game partner is going to receive as a benefit. In each game, the decision of the players is influenced by a different probabilistic variable. This variable represents some feature that is exploited by the global behavior to favor the evolution of cooperation. For the kin selection strategy, relationships are exploited, the relation may be genetic, emotional or of any other type but as long as it its closer, cooperation will be more frequent. In the direct reciprocity case, the evolution of cooperation is linked with the probability of one player to play again with the same partner, and this condition is set because players are responding cooperation with cooperation and defection with defection. The feature exploited by the indirect reciprocity case is peer pressure. While more players get to know the actions of a determined player, she will obtain more benefits by cooperating and while fewer players get to know her actions more benefits she will obtain by defecting. Each strategy has a condition for the cooperation to become an evolutionary stable strategy (ESS, cooperators survive), another when cooperation becomes risk-dominant (RD, cooperators are near 1/2 of the population) and other when cooperators are advantageous (AD, cooperators are near 2/3 of the population).

| Strategy | Payoff table | | | ESS | RD | AD | Variables |
|---|---|---|---|---|---|---|---|
| Kin Selection | | C | D | $\dfrac{b}{c} > \dfrac{1}{r}$ | $\dfrac{b}{c} > \dfrac{1}{r}$ | $\dfrac{b}{c} > \dfrac{1}{r}$ | b=benefit c=cost r=relatedness probability |
| | C | (b-c)(1+r) | br-c | | | | |
| | D | b-rc | 0 | | | | |
| Direct Reciprocity | | C | D | $\dfrac{b}{c} > \dfrac{1}{w}$ | $\dfrac{b}{c} > \dfrac{2-w}{w}$ | $\dfrac{b}{c} > \dfrac{3-2w}{w}$ | b=benefit c=cost w=probability of next round |
| | C | (b-c)/(1-w) | -c | | | | |
| | D | b | 0 | | | | |
| Indirect Reciprocity | | C | D | $\dfrac{b}{c} > \dfrac{1}{q}$ | $\dfrac{b}{c} > \dfrac{2-q}{q}$ | $\dfrac{b}{c} > \dfrac{3-2q}{q}$ | b=benefit c=cost q=social acquaintanceship |
| | C | b-c | -c(1-q) | | | | |
| | D | c(1-q) | 0 | | | | |

**Table 1.** Rules for the three game strategies [4].

## 2  The Models

We developed agent-based computational models, using the NetLogo [12] platform, to better understand three strategies for the evolution of cooperation: kin selection, direct reciprocity and indirect reciprocity [4]. Unlike the results shown in [4] these models are focused in cultural evolution (players may choose to cooperate or not under different circumstances instead of been born as players who always cooperate or always defect), therefore, the first characteristic these models share is that they have a constant population. The behavior of the agents is similar to the one of agents described in Skyrm's Matching Pennies model [8]. Agents have a cooperation probability variable (*cp*) that determines their attachment to a cooperator strategy, while this variable gets a greater value the agent will cooperate more frequently. We will use this variable for deciding if the agents will cooperate or not in a particular game. Each agent is able to identify herself as a cooperator or as a defector:

$$\text{If } cp(p_i) > 0.5 \text{ then } c_i = p_i \text{ and } d_i = 0$$

$$\text{If } cp(p_i) \leq 0.5 \text{ then } d_i = p_i \text{ and } c_i = 0$$

Where $p_i$ is the i$^{th}$ agent or player, $c_i$ is the i$^{th}$ cooperator and $d_i$ is the i$^{th}$ defector.

As a consequence of this characteristic, a partition of the population can be made. The *initial proportions of cooperative versus defective* agents in the population (*ipc* and *ipd*) are relevant parameters of the model although they can be reduced to one as *ipc*=1-*ipd*. In order to identify agents as cooperators or defectors at the beginning of the simulations we give them an initial *cp*, thus, *initial cp of defectors (iicpd)* and *initial cp of cooperators (icpc)* are two more parameters. The agents move through a defined two-dimensional space. It has been seen [3] that in such cases population density is a key determinant of the dynamics, so the number of players (*population*) interacting in the defined space is another important parameter of the model.

Agents interact following the payoff tables corresponding to each strategy studied. As can be seen in Table 1, the three strategies share two variables: the benefit (*b*) one agent gives to another while cooperating and the cost (*c*) paid by the cooperator while cooperating. With the purpose of keeping track of the player's decisions, each one of them is assigned a *fitness* value from where the cost will be subtracted and the benefits will be added.

Appendix A shows the instructions executed by the agents during each iteration of the simulations. The behaviors of each game are detailed in appendix B.

### 2.1 Tune cooperation probability

Agents adapt to the environment by modifying whether they are going to cooperate or defect in the next round. We implemented several exogenous tuning algorithms based on the literature [8][9][6][7][11] with no satisfactory results. We developed two self-organizing tuning alternatives and were tested with very good results:

- **Selfish fitness**: the agents only take their own *fitness* value as a parameter, unlike other algorithms that take into account the values of all the other players or values of the game partners. If the player cooperates or defects, and her *fitness* increases, then the agent increases or decreases her *cp* respectively. If the player cooperates or defects, and her *fitness* is reduced, then the agent decreases or increases her *cp* respectively. And if there is no change in the agent's fitness, then there is no change in her cp.

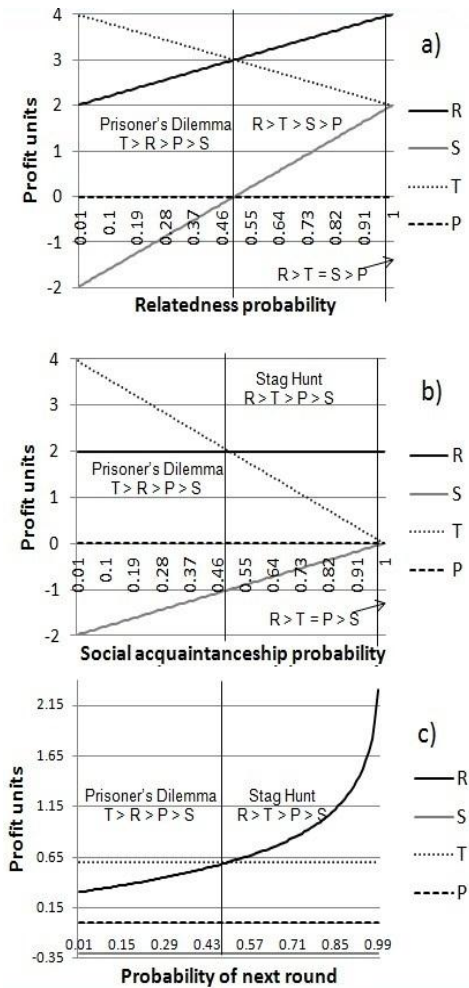

- **Selfish profit**: this is very similar to the previous one, but instead of comparing fitness values (sum of all profits obtained), agents act based in the comparison of the profit of their last game and the profit of the current round. Thus, if the player cooperates or defects and the actual profit is greater than the last one, then the agent increases or decreases her *cp*. If the player cooperates or defects and the actual profit is lower than the last one, then the agent decreases or increases her *cp*. Finally if the last profit is equal to the actual profit, then there is no change in the *cp* value.

## 3 Methods

Before general results were obtained, an analysis of the variables was conducted. A primary goal of this analysis was to determine the impact of the parameters on the dynamics of the models. A secondary goal was to obtain parameter values that best exemplify the model's behavior. Detailed information about the experiments may be found in appendix C.

**Fig. 1.** Graphics of the payoff table behaviors for x's values in the range [0,1] for a) Kin Selection, b) Indirect Reciprocity and c) Direct Reciprocity (DR is in logarithmic scale).

## 4 Results

Interesting results can be derived from a careful analysis of the graphics of the payoff table behavior as the corresponding probability takes values within the range [0,1] (Figure 1). As *x* has a higher value, the agents move to different games. Using the same notations as in [6] that designates R (reward) when two agents cooperate, P (punishment) when two players defect, T (temptation) when one player defects and the other cooperates and S (sucker´s payoff) when one player cooperates and the other defects; the move from game to game is described in appendix D. The values of *b* and *c* also determine the transition value of *x* that takes the agents from one game to another. It is important to notice that the agents respond to these transitions by interacting with the others and adapt by using only simple local rules with only two basic requirements: 1) the agents must know how to distinguish that an amount is bigger than other and 2) the agents must have the notion of more is better. Within the strategies studied, other requirements are implicitly given for each case and are described in the appendix E.

### 4.1 ESS, RD and AD conditions

It was shown in Table 1, when the *x* variable reaches certain value, ESS, RD and AD behaviors emerge. Figure 2 shows the characteristic behavior obtained by the analysis described in the behavior part of the Methods section; in it, these expected behaviors are not noticeable. Analyzing the model variables, we could find that these conditions are preserved, but the effect can be observed by setting the starting population of cooperators as 2/3, 1/2 and 1/3 of the total population for ESS, RD and AD respectively and obtaining a graphic similar to Figure 2. This result is shown in appendix F along with the impact of all the parameters to each strategy.

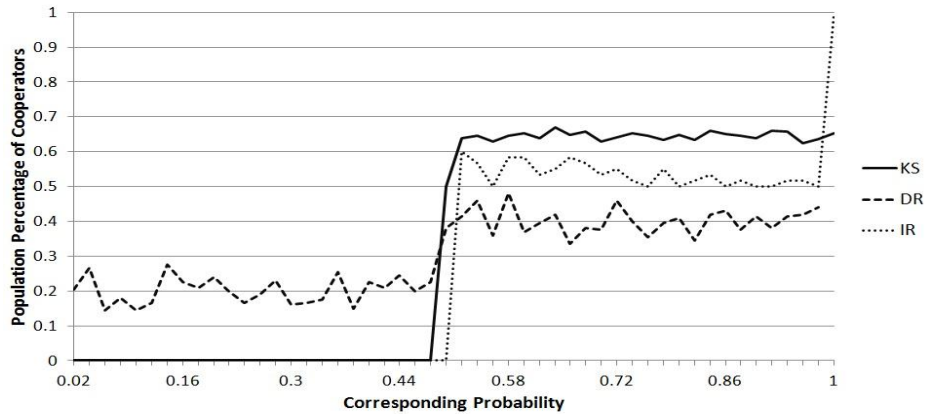

**Fig. 2.** Percentage of cooperating population while the corresponding probability is varied.

### 4.2 Self-organization

Self-organization seems to be a key element for the cultural evolution of cooperation. There are several studies [8][9][6][7][11] in which the agents just can choose between cooperate or defect. In the models presented here, the players can choose the degree of cooperation that they consider the best. Another important difference between our models and the others [6][7] is the inclusion of a probabilistic variable so that agents choose the degree of cooperation by comparing their actual state with a past one instead of comparing their state with the state of others. This is important because each agent may be seen as an individual system adaptable to different environments or contexts using the information given by the interactions with other similar, not necessarily equal, systems.

## 5      Conclusions

Three agent-based computational models for the study of the evolution of cooperation under cultural propagation were described. It was shown that their behavior is the result of transitions between games defined by Game Theory. The transitions are consequences of the structure determined by the payoff matrixes of the three strategies studied. Each of these strategies abstracts well-known real behaviors [4]; hence the importance of creating computational models that let us experiment exhaustively different circumstances for these phenomena, a difficult task with living organisms. The impact of the parameters in each model was analyzed to better understand how to manipulate the models and adapt them for more specific studies, such as social phenomena (e.g. bulling, market formation, and migration), traffic problems (e.g. signaling issues, conflicting use of roads) and biological processes (e.g. interactions between organisms and populations).

# Appendixes

## A: General behavior of the agents.

```
program Iteration (p_i)
{It must be guaranteed that each agent has followed the
last instruction, then the next one may be executed}
var   cooperate(p_i): Boolean ;
var   p_j: Agent ;
begin
  1 rotate left (random (50));
  2 rotate right (random(50));
  3 move forward number of steps(1);
  4 find neighbors at distance (1);
  5 p_j:= engage ;{this procedure are going to return the
  game partner of agent p_i}
  6 play ;
  7 get consequences ;
  8 tune cooperation probability ;
end.
```

The lines 1-3 let the agents set a straight line direction but with some noise. In line 4, the agents look for other agents near them within one cell distance. The engage instruction prompts the agents to choose one neighbor (if any). Once a pair of agent engages in a game they became ineligible for other agents. The procedures in lines 6-7 contain details of each strategy.

## B: Games behavior.

### Kin selection

Under this strategy the agents, engaged in a game, participate in a naive way supposing that they are involved in some kind of relationship. Although a player may be a cooperator, there are going to be conditions under which she will not cooperate. For example, she might not be in the mood or she isn't fast enough to do the action required by the other agent. The random decision based on the variable *cp* permits to simulate this and also how an agent overlooks these conditions while her cooperator attitude gets stronger (*cp* gets higher values).

```
program play kin selection
var   decision: Real;
begin
  decision:= random (0,1);
  if ( decision <= cp(p_i) )
```

```
      cooperate(pᵢ):= true ;
   else
      cooperate(pᵢ):= false ;
end
```

Once the players have simultaneously defined their actions, they move to the get consequences phase in which they add or subtract units from their *fitness* following the upper section of Table 1.

The tune cooperation probability phase is explained later in its own section.

### Direct Reciprocity

In [5] a detailed explanation of the agents is given. There are two types of agents: a) agents following an always defect strategy (defectors) and b) agents following tit-for-tat strategy (cooperators) that will start with cooperation. Technically, a Boolean array may be added to each agent for the tit-for-tat strategy implementation so, on the array of a player $p_i$ the last action of an agent $p_j$ will be recorded. To analyze the cultural propagation of cooperation, adjustments to these rules were made and the behavior is the following:

```
program play direct reciprocity
var:  decision: Real ;
begin
  decision:= random (0,1) ;
  if (decision <= cp(pi))
  begin
    if (array[j] == true)
      cooperate(pᵢ):= true ;
    else
      cooperate(pᵢ):= false ;
  end
  else
  begin
    cooperate(pᵢ):= false ;
  end
end
```

For the get consequences phase, once the players have simultaneously defined their actions they add or subtract units form their *fitness* following the middle section of Table 1.

### Indirect reciprocity

The development of the games under this strategy is also specified in [5]. Here the $q$ value has a more important role. Before making a decision, each player has the probability of knowing if his game partner is a defector or a cooperator. A defector will never cooperate, while a cooperator will always cooperate if his partner is a cooperator and will not cooperate if the partner is a defector. A cooperator will cooperate with a defector with a probability 1-$q$, simulating the fact that the cooperator couldn't recognize the other as a defector. To study the cultural evolution of cooperation, the game is structured like the next code:

```
program play indirect reciprocity
var: decision, recognition: Real ;
begin
  decision:= random (0,1) ;
  recognition:= random (0,1) ;
  if (p_i == c_i and p_j == c_j) {if both players are coopera-
  tors}
  begin
    if (recognition <= (1 - q)) {if p_i don't recognize p_j
    as a cooperator}
    begin
      if (decision <= cp(p_i))
        cooperate(p_i):= true ;
      else
        cooperate(p_i):= false ;
    end
    else {if p_i recognize p_j as a cooperator}
      cooperate(p_i)=true ;
  end
  else
  begin
    if (p_i == c_i and p_j == d_j) {if p_j isn't a cooperator}
    begin
      if (recognition <= (1 - q)) {if p_i don't recognize
      p_j as a defector}
      begin
        if (decision <= cp(p_i))
          cooperate(p_i):= true ;
        else
          cooperate(p_i):= false ;
      end
      else {if p_i recognize p_j as a defector}
        cooperate(p_i):= false ;
    end
    else
    begin
      if (p_i == d_i) {if p_i is a defector}
      begin
        if (decision <= cp(p_i))
```

```
            cooperate(p_i):= true ;
         else
            cooperate(p_i):= false ;
       end
     end
   end
end
```

When the players have simultaneously defined their actions, the behavior of the get consequences phase is to add or subtract units form their *fitness* following the lower part of Table 1.

## C: Detailed methodology.

The size of the space in which agents interacted was thirteen by thirteen cells with 1:1 agent-cell proportion and each agent step was proportioned to 1:1 cell length. The space has no boundaries so it has a toroidal shape.

Each strategy was treated with similar analyzes with few differences in sampling. The analyzes were constructed with the idea of studying the global system in stable conditions thus it is similar to the one described in [3] and consist of simulations running 100,000 iterations outputting the average percentage of cooperators in the population. The averages were taken for the last 5,000 iterations. A simulation was made for each possible configuration described in the following sections:

- **Payoff table behavior**: as it can be seen in Table 1, each strategy is dependent on three variables: $b$, $c$, and depending on the strategy, $r$, $w$, or $q$. Variables $b$ and $c$ were fixed $b=4$ and $c=2$ arbitrarily, while $r$, $w$ or $q$ (that from now on will be denoted as $x$) took values from the range [0.01 , 1] for kin selection (KS) and indirect reciprocity (IR) strategies and from [0.01 , 0.99] for direct reciprocity (DR). In each case, samples were taken every 0.01 steps in the $x$'s range. $w=1$ was omitted to avoid division by zero.

- **Tuning criterion (TC)**: as explained in the Model section, after each game the agents try to adapt their behavior to the environment (tune cooperation probability). There are two criteria to do so: one based on the agent's fitness (sf) and another one based on the agent's profit (sp). For this part of the analysis, each parameter was arbitrarily fixed to the values described in Table 2.

|    | B | C | icpc | icpd | population | ipc | ipd | X's range | Sampling interval | TC |
|----|---|---|------|------|------------|-----|-----|-----------|-------------------|------|
| KS | 4 | 2 | 0.65 | 0.35 | 60         | 0.5 | 0.5 | [0.01- 1] | 0.01              | [sf, sp] |

| | | | | | | | | | |
|---|---|---|---|---|---|---|---|---|---|
| DR | 4 | 2 | 0.65 | 0.35 | 20 | 0.5 | 0.5 | [0.01- 0.99] | 0.01 | [sf, sp] |
| IR | 4 | 2 | 0.65 | 0.35 | 60 | 0.5 | 0.5 | [0.01- 1] | 0.02 | [sf, sp] |

**Table 2.** Parameter's settings for experiments in Tuning criterion section.

– **Initial probabilities**: as mentioned while describing the model, each agent begins with some probability for cooperation. Starting from this section the results from previous ones are taken into account, thus, the parameters used for this experiments are detailed in Table 3.

| | B | C | icpc | icpd | population | ipc | ipd | X's range | Sampling interval | TC |
|---|---|---|---|---|---|---|---|---|---|---|
| KS | 4 | 2 | [0.65, 0.75, 0.85, 0.95, 0.99] | [0.01, 0.05, 0.15, 0.25, 0.35] | 60 | 0.5 | 0.5 | [0.01- 1] | 0.05 | **sf** |
| DR | 4 | 2 | [0.65, 0.75, 0.85, 0.95, 0.99] | [0.01, 0.05, 0.15, 0.25, 0.35] | 20 | 0.5 | 0.5 | [0.01- 0.99] | 0.05 | **sp** |
| IR | 4 | 2 | [0.5, 0.51, 0.55, 0.65, 0.75, 0.85, 0.95, 0.99] | [0.01, 0.05, 0.15, 0.25, 0.35, 0.45, 0.49, 0.5] | 60 | 0.5 | 0.5 | [0.01- 1] | 0.05 | **sp** |

**Table 3.** Parameter's settings for the experiments of the Initial probability section. Values in bold font represent results of previous sections.

– **Population**: in this part the impact of the number of players was explored. It was expected that there would be no impact for KS and IR but, in DR the population should be an important issue. The parameters used are described in Table 4.

|    | B | C | icpc | icpd | population | ipc | ipd | X's range | Sampling interval | TC |
|----|---|---|------|------|------------|-----|-----|-----------|-------------------|-----|
| KS | 4 | 2 | **0.65** | **0.35** | [10, 20, 30, 40, 50, 60, 70, 80, 90, 100] | 0.5 | 0.5 | [0.01- 1] | 0.04 | **sf** |
| DR | 4 | 2 | **0.65** | **0.35** | [2, 4, 6, 8, 10, 20, 30, 40, 50, 60, 70, 80, 90, 100] | 0.5 | 0.5 | [0.01- 0.99] | 0.04 | **sp** |
| IR | 4 | 2 | **0.98** | **0.45** | [10, 20, 30, 40, 50, 60, 70, 80, 90, 100] | 0.5 | 0.5 | [0.01- 1] | 0.04 | **sp** |

**Table 4.** Parameter's settings for the experiments of the Popultaion section. Values in bold font represent results of previous sections.

**Robustness**: as the name would suggest, the idea behind this test was to know the robustness of the strategy by varying the *initial proportion of cooperators*. The initial proportions were [0, 0.333, 0.5, 0.666, 1], complementary *initial proportion of defectors* were fixed to get the *population* discussed in the population section. The rest of the parameters were fixed as Table 5 details.

|    | B | C | icpc | icpd | population | ipc | ipd | X's range | Sampling interval | TC |
|----|---|---|------|------|------------|-----|-----|-----------|-------------------|-----|
| KS | 4 | 2 | **0.65** | **0.35** | 60 | [0, 0.333, 0.5, | [1, 0.666, 0.5, | [0.01- 1] | 0.02 | **sf** |

| | | | | | 0.666, 1] | 0.666, 0] | | | |
|---|---|---|---|---|---|---|---|---|---|
| DR | 4 | 2 | **0.65** | **0.35** | **20** | [0, 0.333, 0.5, 0.666, 1] | [1, 0.666, 0.5, 0.666, 0] | [0.01- 0.99] | 0.02 | **sp** |
| IR | 4 | 2 | **0.98** | **0.45** | **60** | [0, 0.333, 0.5, 0.666, 1] | [1, 0.666, 0.5, 0.666, 0] | [0.01- 1] | 0.02 | **sp** |

**Table 5.** Parameter's settings for the experiments of the Robustness section. Values in bold font represent results of previous sections.

– **Behavior**: finally average behaviors were measured. The parameters were fixed as showed in Table 6 and the results are the averages of the outputs of 10 simulations for every sample interval.

| | B | C | icpc | icpd | population | ipc | ipd | X's range | Sampling interval | TC |
|---|---|---|---|---|---|---|---|---|---|---|
| KS | 4 | 2 | **0.65** | **0.35** | **60** | 0.5 | 0.5 | [0.01- 1] | 0.02 | **sf** |
| DR | 4 | 2 | **0.65** | **0.35** | **20** | 1 | 0 | [0.01- 0.99] | 0.02 | **sp** |
| IR | 4 | 2 | **0.98** | **0.45** | **60** | 0.5 | 0.5 | [0.01- 1] | 0.02 | **sp** |

**Table 6.** Parameter's settings for the experiments of the Behavior section. Values in bold font represent results of previous sections.

## D: Game transitions

| Strategy | Graph section | Parameters | Game | Tension(s) included in each dilemma |
|---|---|---|---|---|
| Kin selection | $\dfrac{b}{c} < \dfrac{1}{r}$ | T > R > P > S | Prisoner's Dilemma | Players prefer T to R. Players prefer P to S. |
| Kin selection | $\dfrac{b}{c} > \dfrac{1}{r}$ | R > T > S > P | Unidentified game | Cooperators always win. Players prefer T to S. |
| Kin selection | $r = 1$ | R > T = S > P | Unidentified game | Cooperators always win. There are no preference between T and S. |
| Direct reciprocity | $\dfrac{b}{c} < \dfrac{1}{w}$ | T > R > P > S | Prisoner's Dilemma | Players prefer T to R. Players prefer P to S. |
| Direct reciprocity | $\dfrac{b}{c} > \dfrac{1}{w}$ | R > T > P > S | Stag Hunt | Players prefer P to S |
| Direct reciprocity | $w \approx 1$ | R > T > P > S | Stag Hunt | Players prefer P to S |
| Indirect reciprocity | $\dfrac{b}{c} < \dfrac{1}{q}$ | T > R > P > S | Prisoner's Dilemma | Players prefer T to R. Players prefer P to S. |
| Indirect reciprocity | $\dfrac{b}{c} > \dfrac{1}{q}$ | R > T > P > S | Stag Hunt | Players prefer P to S |
| Indirect reciprocity | $q = 1$ | R > T = P = S | Unidentified game | Only when both players cooperate they win. |

**Table 7.** Comparison of the game transitions using a format similar to the one used in [6].

## E: Implicit requirements for each game.

### Kin selection

The additional condition is: agents must be able to determining the probability of genetic relatedness to other agents. In nature this is frequently observed, for example, a tiger is able of realize that it is in the territory of another tiger by the smell. Independently of the actual mechanisms used by living systems for determining genetic relatedness, we have found that this strategy is very robust. Independently of the initial proportion of cooperators in the population, if $b/c > 1/r$ then, cooperators and defectors can coexist. A particular question may arise: if kin selection is so robust for the evolution of cooperation, why cooperative behaviors do not always occur among individuals of the same species, such as bulling between students or the same tiger of the last example fighting for territory with another tiger? One possible answer is that the ability to estimate relatedness depends on the individuals, but the perceptions of $b$ and $c$ are codependent of the environment and of the subjective perception of individuals. A clear and sad example of this is: the polar bears, when a female polar bear has cubs and there have been no food for several days, the mother can commit cannibalism. In this case the genetic relatedness perception remains the same (r=1) but the lack of food increases perception of the cost value for the mother and a deplorable non-cooperative consequence occurs.

### Direct reciprocity

At a first glance, the additional condition of this case seems to be the ability of agents to know the probability that they have of playing again with another particular agent. Analyzing this strategy, it may be noticed that the results of this case were those with the highest error rate compared with the expected behavior. The most reasonable explanation is that, somehow, the agents notice the difference between the given $w$ (set as a parameter) and the real one (given by the number of players in the area). One possibility is that agents attend so much to the given $w$ with wrong results that they stop paying attention to it, although this requires further analysis.

### Indirect reciprocity

The additional condition here is the ability of the agents to estimate in which proportion they actions are known by the rest of the population. This ability is only seen in organisms with complex social interactions. The analysis conducted showed that the global behavior of this case is sensitive to the initial probabilities for cooperation of the players; it was necessary to fix *initial cp of cooperator*=0.99 and *initial cp of defectors*=0.5 for the model to show the expected behavior (ESS behavior under $b/c > 1/x$ conditions). When *icpc* set lower or *icpd* is set higher, the point in the $x$'s range in which the cooperating population may survive is displaced to the right, thus, more players knowing the actions of other players are required for cooperation to evolve. This may be understood as the effect of a cultural stance to a particular subject. For example, corruption: if the population has cultural aspects that favor corruption, then the initial probability of cooperators and defectors will have a low value.

## F: Impact of the parameters to each strategy

| Parameter | Effect in strategy | | |
|---|---|---|---|
| | Kin Selection | Direct Reciprocity | Indirect Reciprocity |
| Selfish fitness | Interaction between players and parameters allow the emergency of ESS, RD and AD behaviors. | No distinguishable patterns can be observed. | Interaction between players and parameters allow the emergency of ESS, RD and AD behaviors. |
| Selfish profit | If $b/c < 1/r$ all players are defector. If $b/c > 1/r$ all players are cooperators. In both cases all the other parameters are irrelevant. | Interaction between players and parameters allow the emergency of distinguishable patterns for ESS, RD and AD behaviors. | Interaction between players and parameters allow the emergency of ESS, RD and AD behaviors but later in $x$'s range when compared with the results of selfish fitness. |
| Benefit (b) | Sets highest payoff that R can get. Sets starting value for T in the $x$'s range. Along with $x$ and $c$ sets the value for R. | Sets the value for T in the $x$'s range. Along with x and c sets the value for R. | Sets the starting point for T in the $x$'s range. Along with $c$ sets the value for R. |
| Cost (c) | Sets starting point for T. Sets starting and final points for S in the x's range. Along with $x$ and $b$ sets the value for R. | Sets the value of S. Sets the starting point for R in the x's range. Along with $x$ and $b$ sets the value for R. | Sets the starting point for S in the x's range. Along with $b$ sets the value for R. |
| Initial probability of cooperators | No significant impact. | Noise generating variable; with higher values, the behavior becomes noisier in the $x$'s range points under $b/c < 1/x$ | Noise generating variable; with lower values, the behavior becomes noisier. |
| Initial probability of defectors | Noise generating variable; with lower values, the behavior becomes noisier. | No significant impact | Noise generating variable; with lower values, the behavior becomes noisier. |
| Population | No significant impact | Noise generating variable; with higher values the behavior becomes noisier. | Noise generating variable; with lower values, the behavior becomes noisier. |
| Proportion of initial cooperators | No significant impact | With higher values the final proportion increases. | With higher values the final proportion increases. |

**Table 8.** Comparison of the impact of the parameters in each strategy.